\documentclass[pra,twocolumn,showpacs,floatfix]{revtex4}

\usepackage{amsmath}
\usepackage{times}
\usepackage{graphicx}
\usepackage[dvips]{color}

\begin{document}

\title{Optical tuning of the scattering length of cold alkaline earth atoms}
\author{R. Ciury\l{}o$^{\dagger\S}$, E. Tiesinga$^{\dagger}$, P. S. Julienne$^{\dagger}$}
\affiliation{$\dagger$ Atomic Physics Division, National Institute of Standards and Technology,
100 Bureau Drive, Stop 8423, Gaithersburg, Maryland 20899-8423, USA.\\
$\S$ Instytut Fizyki, Uniwersytet Miko\l{}aja Kopernika,
ul. Grudzi\c{a}dzka 5/7, 87--100 Toru\'n, Poland.}
\date{\today}

\begin{abstract}
It is possible to tune the scattering length for the collision of ultra-cold  $^{1}$S$_{0}$ ground
state alkaline-earth atoms using an optical Feshbach resonance.  This is achieved with a
laser far detuned from an excited molecular level
near the frequency of the atomic intercombination $^{1}S_{0}$--$^{3}P_{1}$
transition.  Simple resonant scattering theory, illustrated by the example of  $^{40}$Ca,
allows an estimate of the magnitude of the effect.  Unlike alkali metal species,
large changes of the scattering length are possible while atom
loss remains small, because of the very narrow line width of the molecular photoassociation
transition.  This raises prospects for control of atomic
interactions for a system without magnetically tunable Feshbach resonance levels.
\end{abstract}

\pacs{34.50.Rk, 34.10.+x, 32.80.Pj}

\maketitle

In recent years the ability to change the interaction between ultra-cold colliding
atoms has opened the way for new and exciting experiments
with ultra-cold atomic gases.  The observation of a Bose-Einstein
condensate (BEC) in atomic cesium \cite{weber03} would have
been impossible without the ability to change the interaction from
being attractive to being repulsive. More impressively, time-varying
interactions have allowed the creation of condensates of two-atom molecules
starting from an atomic Bose condensate\cite{greiner03,jochim03}.
Most recently, the observation of the condensation of pairs of
fermionic \cite{regal04} atoms has started investigations into the
so-called BEC-BCS crossover\cite{williams04}, where BCS is the abbreviation
for the Bardeen-Cooper-Schrieffer phase transition in a fermionic gas\cite{bardeen57}.

The key to these developments has been the ability to change
the interaction between atoms
by a magnetically-tuned Feshbach resonance \cite{tiesinga93}.
Theoretical discussion of the properties of these resonances
can be found in Refs.~\cite{mies00,raoult04,marcelis04,goral04}.
The interaction between the atoms at ultra-cold temperature can be characterized
by a single parameter, the scattering length\cite{weiner99,burnett02}, which can be
controlled in sign and magnitude using these resonances.

Another way to change the scattering length of two colliding atoms
is to optically couple the ground scattering state with an excited
bound state\cite{fedichev96}.  These optical Feshbach resonances are
theoretically analyzed in Refs.~\cite{bohn97,bohn99} and implemented
experimentally in Refs.~\cite{fatemi00,theis04}. The recent experiment of Theis {\it et
al}. \cite{theis04} with $^{87}$Rb atoms showed, however, that a significant change of the
scattering length is accompanied with substantial loss of atoms.  The same is true
if a two-color Raman process is used~\cite{Thalhammer04}.

In this paper we discuss the optical tuning of scattering lengths in
ultra-cold alkaline-earth atom vapors. To do so we will assume that the laser
is far detuned from excited molecular states near the intercombination
transition, $^{1}S_{0}$--$^{3}P_{1}$, as recently analyzed in
Ref.~\cite{ciurylo04}. We show that significant changes of the scattering
strength can be achieved without the excessive atom loss
that plagues experiments with alkali-metal gases\cite{theis04}.
Prospects for an optically tuned scattering length in ultra-cold
alkaline-earth vapors seems to be particularly attractive, since
magnetically tuned Feshbach resonances do not occur between
the isotopes of ground-state $^{1}S_{0}$ alkaline-earth atoms with zero nuclear spin.
Several alkaline earth metal atoms also have isotopes with nonzero integer
or half-integer nuclear spin.  Optical methods could be used to tune their
scattering lengths, although we will not consider these specifically in this
paper.  Optical Feshbach control can be also applied to other atomic
systems having a similar electronic structure. The recent
Bose-Einstein condensation of ytterbium \cite{takasu03} makes this
system especially interesting.

The theoretical description of optical Feshbach resonances and the
closely related photoassociation (PA) process is well established
\cite{fedichev96,bohn97,bohn99}.  The elastic and inelastic scattering rates
due to a single photoassociation resonance level depends on (1) the natural line
width $\Gamma_{e,\rm nat} \approx  2\Gamma_A$ of the excited molecular level
 $e$, (2) the stimulated width $\Gamma_{eg}(\varepsilon_{r})$ that couples the excited
level to the $s$-wave collision of the ground state $g$ at
relative kinetic energy $\varepsilon_{r}$,
and (3) the detuning $\Delta-\Delta_{e}$ from optical resonance.  Here,
following the notation of Ref.~\cite{ciurylo04} for PA near the $^1$S$_0$-$^3$P$_1$
intercombination line of a Group II atomic species,  $\Gamma_A$ is the natural decay
width of the atomic transition, $\Delta=\hbar\omega-E_A$
 $\Delta_{e}=E_e-E_A$, where  $E_e$ is the energy of an isolated excited
molecular bound level, $E_A$ is the energy of the $^3$P$_1$ atom,
and $\omega$ is the frequency of the light driving the transition.  The stimulated
width is
\begin{equation}
  \Gamma_{eg}(\varepsilon_{r}) = 2 \pi |\langle e | V_\mathrm{las} | g \rangle |^2
   = \Gamma_A \frac{3}{4\pi}\frac{I\lambda_A^3}{c} f_\mathrm{rot} f_\mathrm{FC} \,,
   \label{StimGamma}
\end{equation}
where $V_\mathrm{las}$ is the optical coupling proportional
to the laser intensity $I$, $\lambda_A$ is the wavelength of the atomic transition,
$c$ is the speed of light, $f_\mathrm{rot}$ is a dimensionless rotational line strength factor of
order unity, and $f_\mathrm{FC}$ is the Franck-Condon factor per unit energy
for the free-bound PA transition:
\begin{equation}
  f_\mathrm{FC} = \left | \int_0^\infty F_e(R) F_g(\varepsilon_{r},R) dR \right |^2 \,.
  \label{FCF}
\end{equation}
Here, $F_e$ is the unit normalized excited state wave vibrational
function and $F_g(\varepsilon_{r},R)$ is the energy normalized
ground state scattering wave function.  The low energy $s$-wave
form of the latter is $(2\mu/\pi \hbar^2 k_r)^{1/2}
\sin(k_r(R-a_\mathrm{bg}))$ at large $R$ , where $\mu$ is reduced mass of the atom
pair, $\hbar$ is  Planck's constant divided by $2\pi$,
$k_r=\sqrt{2\mu\varepsilon_{r}}/\hbar$, and $a_\mathrm{bg}$ is the
ground state $s$-wave scattering length in the absence of light.
It should be noted that the details of the molecule structure are
hidden in $\Gamma_{eg}(\varepsilon_{r})$, $\Delta_{e}$, and
$a_\mathrm{bg}$.

The rate constant for inelastic collisions that lead to atom loss is typically large
when the detuning from molecular resonance is small. Consequently, we will
only consider the case of unsaturated transitions at large detuning, defined by the condition
\begin{equation}
 |\Delta-\Delta_{e}|\gg\Gamma_{e,\rm nat}+\Gamma_{eg}(\varepsilon_{r}) \,.
 \label{LargeDet}
\end{equation}
We also require that $|\Delta-\Delta_{e}|$  be much larger than
other contributions to the width of the photoassociation line such as the
thermal width, Doppler width \cite{ciurylo04}, light induced shift
 \cite{bohn99,simoni02}, and the mean-field shift in the case of BEC \cite{killian00}.

The theoretical description can be framed using the definition of
a complex scattering length ${\cal A}$ based on the elastic scattering
$S$-matrix element as $k_r \to 0$~\cite{bohn97}
\begin{equation}
\label{Sgg}
    S_{gg} = \exp ( -2 i {\cal A} k_r )\,.
\end{equation}
The length ${\cal A}$ is complex and in the presence of light can be written as
\begin{equation}
\label{cAscat}
{\cal A}(\Delta,I)=a_{\rm bg}+a_{\rm opt}(\Delta,I)-ib_{\rm opt}(\Delta,I) \,,
\end{equation}
where the dependence on $\Delta$ and $I$ are made explicit.  The
optically induced $a_{\rm opt}(\Delta,I)$ and $b_{\rm
opt}(\Delta,I)$ vanish for $I=0$ and are linear in $I$ for the
limit in Eq.~(\ref{LargeDet}). The length
$a_{\rm scat}(\Delta,I)=a_{\rm bg}+a_{\rm opt}(\Delta,I)$ is interpreted
as the usual scattering length, and $b_{\rm opt}$ is related to
the atom loss rate coefficient $ K(\Delta,I)$ determined by the unitary
$S$-matrix:
\begin{equation}
\label{Kinel}
\lim_{k_r \to 0} K(\Delta,I)=
   \lim_{k_r \to 0} \frac{\pi\hbar}{\mu k_r} ( 1 - |S_{gg}|^2) =
   \frac{4\pi\hbar}{\mu} b_{\rm opt}
    \,.
\end{equation}
The real and imaginary parts of ${\cal A}$ are directly related to the mean field
energy and the lifetime of a Bose-Einstein condensate.
For the case of a condensate at density $n$, the requirement
$|a_{\rm scat}| \gg b_{\rm opt}$ ensures that the mean field energy per atom pair
$4\pi\hbar^2 a_{\rm scat} n/\mu$ is large compared to the decay width
$\hbar Kn = 4\pi\hbar^2 b_{\rm opt} n/\mu$.  The time scale for decay
is $(Kn)^{-1}$ for a condensate and $(2Kn)^{-1}$ for a noncondensed thermal gas ~\cite{stoof89}.
In the noncondensed thermal gas $\dot{n}=-2Kn^{2}$ if other processes are neglected.

Given the large detuning condition of Eq.~(\ref{LargeDet}), the expressions in Refs.~\cite{bohn97,bohn99} reduce to
\begin{eqnarray}
\label{Aopt}
a_{\rm opt}(\Delta,I)&=& \frac{1}{2 k_{r}}
  \frac{\Gamma_{eg}(\varepsilon_{r})}{\Delta-\Delta_{e}}
  = a_{\rm bg} \frac{\delta_{eg}}{\Delta-\Delta_{e}} \,, \\
\label{Ascat}
a_{\rm scat}(\Delta,I)&=& a_{\rm bg}
\left( 1 + \frac{\delta_{eg}}{\Delta-\Delta_{e}} \right) \\
\label{Bopt}
b_{\rm opt}(\Delta,I)&=&\frac{1}{2}a_{\rm opt}(\Delta,I)
\frac{\Gamma_{e,\rm nat}}{\Delta-\Delta_{e}} \,,
\end{eqnarray}
where $a_{\rm bg}$ is the background scattering length defined previously and
\begin{eqnarray}
\label{Deg}
\delta_{eg}&=&\frac{\Gamma_{eg}(\varepsilon_{r})}{2k_{r}a_{\rm bg}} \,.
\end{eqnarray}
Since the threshold properties of low energy scattering ensure that $\Gamma_{eg}(\varepsilon_{r})\propto k_r$
as $\varepsilon_r\to0$, we see that $\delta_{eg}$,  $a_{\rm opt}(\Delta,I)$,
and $b_{\rm opt}(\Delta,I)$ are independent of collision
energy for $\varepsilon_r\to0$.   Equation~(\ref{Ascat}) shows that the change in scattering
length for an optically induced Feshbach resonance has the same form as
that for a magnetically tunable Feshbach resonance when the width $\delta_{eg}$
is used.  With our definitions, $\Delta-\Delta_{e} >0$ corresponds to blue detuning
and a positive change in scattering length.

For optical Feshbach resonances it is also convenient to express
the detuning dependence in terms of the natural linewidth, namely
\begin{eqnarray}
\label{ALopt}
a_{\rm opt}(\Delta,I)&=& l_{\rm opt}
  \frac{\Gamma_{e,\rm nat}}{\Delta-\Delta_{e}}  \,, \\
\label{BLopt}
b_{\rm opt}(\Delta,I)&=&\frac{1}{2}l_{\rm opt}
\left (\frac{\Gamma_{e,\rm nat}}{\Delta-\Delta_{e}}\right )^2 \,,
\end{eqnarray}
where the ``optical length'' is defined as
\begin{equation}
\label{Lopt}
l_{\rm opt}=\frac{\Gamma_{eg}(\varepsilon_{r})}{2k_{r}\Gamma_{e,\rm nat}}
           =  a_{\rm bg} \frac{\delta_{eg}}{\Gamma_{e,\rm nat}}
\,.
\end{equation}
This length depends on the molecular physics parameters of the ground and
excited states but is independent of collision energy in the low energy threshold
regime and is proportional to the laser intensity $I$, given our large detuning
assumption.  The optical length is the same as the radius
introduced in Eq.~(12) of Ref.~\cite{bohn97}.

In order to make useful changes in the scattering length, the change has to be
large while the losses remain small.  The former criterion
requires that $|a_{\rm opt}| \gg |a_{\rm bg}|$, whereas the latter requires that
$|a_{\rm opt}| \gg b_{\rm opt}$.  Equation~(\ref{Aopt}) shows that the first criterion is satisfied
as long as $|\Delta-\Delta_{e}| \ll \delta_{eg}$, whereas Eq.~(\ref{Bopt})
shows that the second is satisfied if $|\Delta-\Delta_{e}| \gg \Gamma_{e,\rm nat}$.
The condition in Eq.~(\ref{LargeDet}) requires a more stringent condition on
the detuning, which we can state in terms of $l_{\rm opt}$ by combining
Eqs.~(\ref{LargeDet}) and (\ref{Lopt}): $|\Delta-\Delta_{e}| \gg \Gamma_{e,\rm nat}
(1+2k_r l_{\rm opt})$, where we may take $k_r$ typical of a collision energy in
the system. Combining these criteria, they may be stated either in terms of the
detuning:
\begin{equation}
  \delta_{eg} \gg |\Delta-\Delta_{e}| \gg (1+2k_r l_{\rm opt})\Gamma_{e,\rm nat} \,,
  \label{critD}
\end{equation}
or the length parameters:
\begin{equation}
\frac{l_{\rm opt}}{1+2k_r l_{\rm opt}} \gg a_{\rm opt}(\Delta,I)   \gg a_{\rm bg} \,.
\label{critA}
\end{equation}

In general it will be difficult to satisfy these criteria for the case of strongly
allowed molecular transitions with large $\Gamma_{e,\rm nat}$ since they
typically have a relatively small $l_{\rm opt}$ at the large detunings that are necessary.
However, using the example of
Ca atoms, we now will show that these criteria can be satisfied by
an intercombination line transition, for which the needed large detuning
can be achieved close enough to atomic resonance that the Franck-Condon
factor is large enough that $l_{\rm opt}$ is not too small.

The first experimental demonstration of the optical tuning of
the $^{87}$Rb scattering length were presented by Theis {\it et
al}.\cite{theis04}.  Measurements were done with a moderate laser
intensity of about $500\;\rm W/cm^{2}$.  The authors observed a tuning
range of the scattering length of about $200\;a_{0}$ accompanied by a trap
loss coefficient as large as $2\times10^{-10}\;\rm cm^{3}/s$ ($a_{0}=
0.05291772\;{\rm nm}$).  The optical length for this strongly allowed
$^{87}$Rb$_2$ transition is $l_{\rm opt}\approx 100\;a_{0}$, which is relatively
small and comparable to $a_{\rm bg}=103\;a_0$.
Therefore, a significant change of the scattering length, i.e.
$a_{\rm opt}\sim a_{\rm bg}$, can only be induced close to resonance and is
accompanied by a large trap loss.

We have calculated properties of optical Feshbach resonances for calcium
near the intercombination line. The molecular structure and transition
dipole moments are evaluated as in Ref.~\cite{ciurylo04}.
The molecular structure is insufficiently known to quantitatively
predict the absolute positions of the excited bound vibrational levels.
It is necessary to measure these positions experimentally.
Moreover, the background scattering length of the ground state is
not precisely known but is believed to be positive and on the order of
a few hundred $a_0$~\cite{allard03}.
Nevertheless, we can map out the values of $l_{\rm opt}$,
$a_{\rm opt}$, and other Feshbach properties as a function of
the background scattering length, binding energy, and laser detuning
and intensity.

\begin{figure}
\includegraphics[angle=0,width=\columnwidth,clip]{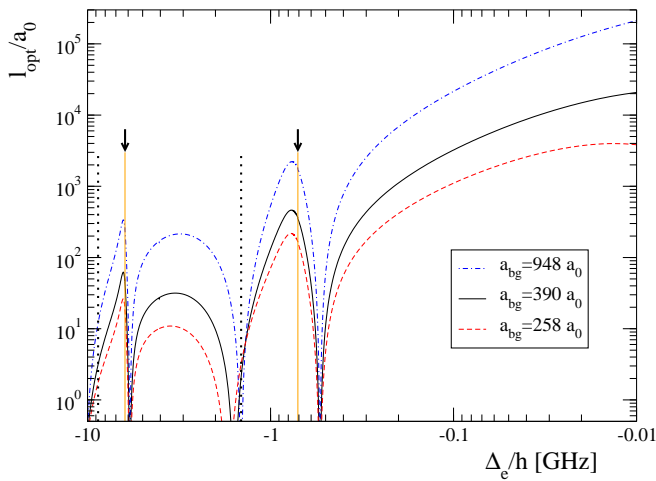}
\caption{(Color online) The optical length $l_{\rm opt}$ calculated
for rovibrational states of the $0_u^+$ band of Ca$_{2}$ as a function
of their line position (binding energy) $\Delta_{e}$ and background
scattering length $a_{\rm bg}$ of the ground state potential. The laser
intensity is 1 W/cm$^2$. The short range of the $0_u^+$ potential is
varied to change $\Delta_e$. The arrows show the position of the
last two unperturbed $1_u$ levels, which are held fixed in the calculation.
The vertical
dotted lines mark the edges of the energy ranges ("bins") within
which one and only one level of $0_u^+$ symmetry will lie, given
some $0_u^+$ potential curve (the first "bin" starts at $\Delta_{e}=0$).
The actual $0_u^+$ and $1_u$ line positions are not known and will need
to be determined experimentally.}
\label{fig1}
\end{figure}

Figure~\ref{fig1} shows  $l_{\rm opt}$ as a function of both
background scattering length $a_{\rm bg}$ and  $\Delta_{e}$ (The
positive binding energy relative to $^3$P$_1+^1$S$_0$ atoms is
$-\Delta_{e}$). Details of the calculation of the stimulated width
$\Gamma_{eg}(\varepsilon_{r})$, necessary for the evaluation of
$l_{\rm opt}$, are described in the figure caption and Ref.~\cite{ciurylo04}.
Figure~\ref{fig1} shows
several maxima and minima in $l_{\rm opt}$.
The optical length at the maxima ranges between $10\;a_0$ and $10^5\;a_0$.
The interference minima to the right of the arrows are due to the mixing of
$0^+_u$ and $1_u$ bound states .  The third minimum is due to vanishing
overlap between excited $0^+_u$ and ground state wave functions. The nature and
properties of these features are discussed in detail in
Ref.~\cite{ciurylo04}.

The figure also shows that the envelop of  $l_{\rm opt}$ (i.e., ignoring oscillations)
increases when $|\Delta_e|$ decreases or $a_{\rm bg}$ increases.
In fact, for a laser intensity of 1 W/cm$^2$
and binding energies on the order of 1 GHz the optical length can
be bigger than $10^3\;a_0$, while it is $0.5\times 10^6\;a_0$ for
$\Delta_{e}\approx -1$ MHz and $a_{\rm bg}\approx 1000\;a_0$.

In order to provide a specific example of $a_{\rm opt}(\Delta,I)$ and
$b_{\rm opt}(\Delta,I)$, we assume a
bound state with $-\Delta_{e}/h=150\;\rm MHz$ and
$a_{\rm bg}=389.8$.  (This optical
resonance corresponds to line number 1 of the $0_u^+$ band as defined in
Ref.~\cite{ciurylo04}.) For this case
$l_{\rm opt}=0.9\times10^{6}\;a_{0}$ at a laser intensity of $500\;\rm
W/cm^{2}$.  This is the same laser intensity as used in Ref. ~\cite{theis04}, but
$l_{\rm opt}$ is four orders of magnitude larger than that for Ref. ~\cite{theis04}.

\begin{figure}
\includegraphics[angle=0,width=\columnwidth,clip]{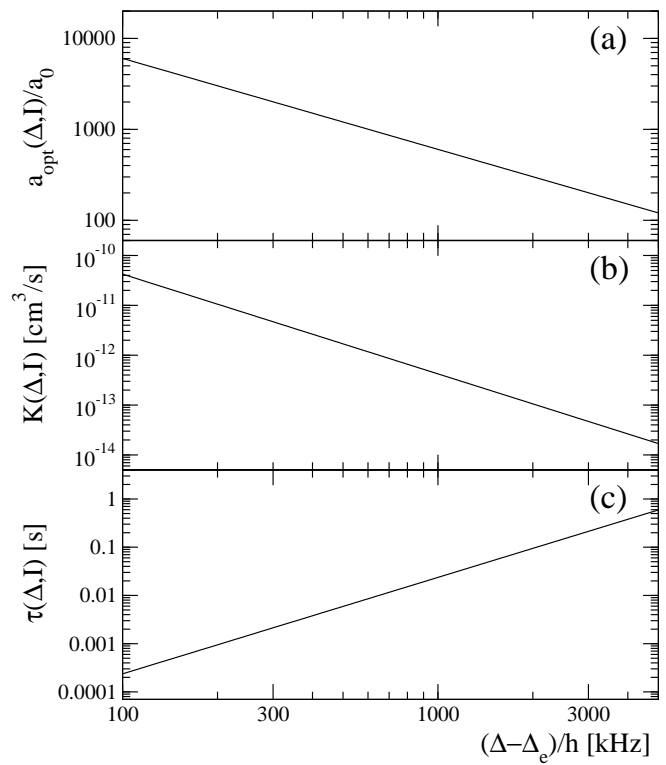}
\caption{(a) The optically induced scattering length $a_{\rm opt}(\Delta,I)$,
(b) the optically induced trap loss rate $K(\Delta,I)$, and
(c) the time scale $\tau(\Delta,I)=[K(\Delta,I)n]^{-1}$ is shown as
a function of the detuning $(\Delta-\Delta_{e})/\hbar$.    The Figure
also applies if the sign of $(\Delta-\Delta_{e})/\hbar$ is reversed,
in which case $a_{\rm opt}(\Delta,I)$ also changes sign.  Results were
obtained for $a_{\rm bg}=389.8$, $\Delta_{e}/h=-150\;\rm MHz$,
$I=500\;\rm W/cm^{2}$, $\Gamma_{e,\rm nat}=0.663\;\rm kHz$,
$l_{\rm opt}=0.9\times10^{6}\;a_{0}$, and $n=10^{14}\;\rm
atom/cm^{3}$.  For collision energy $\varepsilon_{r}/k_{B}=1\;\rm nK$
the stimulated width $\Gamma_{eg}(\varepsilon_{r})/h=18.3\;\rm kHz$
and $k_{r}=1.52\times10^{5}\;a_{0}$.} \label{fig2}
\end{figure}

Figure~\ref{fig2}(a) and (b) show the optically-induced scattering
length $a_{\rm opt}(\Delta,I)$ and loss rate coefficient $K(\Delta,I)$
as a function of blue detuning $\Delta-\Delta_e$ for the parameters
defined above.  The laser detunings shown in the figure are orders of
magnitude larger than the natural linewidth of 0.663 kHz. For these
parameters the stimulated linewidth is 18.2 kHz at a collision energy
of 1 nK, for which $2k_r l_{\rm opt}=27$.  The requirement that
$|\Delta-\Delta_e|/h \gg (1+2k_r l_{\rm opt})\Gamma_{e,\rm nat}/h = 19\;{\rm kHz}$
is easily satisfied. However, for collision energies
on the order of of 1 $\mu$K, $\Gamma_{e,\rm nat}(1+2k_r l_{\rm opt})$
is of the same order as the frequency range in Fig.~\ref{fig2} and our
assumptions are broken. Consequently, for these collision energies we
need a detuning $\Delta-\Delta_e$ that is larger than 5 MHz.

In the figure both $a_{\rm opt}$ and $K(\Delta,I)$ decrease with increasing
detuning. The optically-induced scattering length is $\approx 1200\;a_{0}$
for $(\Delta-\Delta_{e})/h=500\;\rm kHz$, while simultaneously the trap
loss rate is relatively small with $K(\Delta,I)=1.7\times10^{-12}\;\rm
cm^{3}/s$. This loss rate coefficient is two orders of magnitude less than in the
case of rubidium.  The time scale parameter $(Kn)^{-1}\approx 6\;\rm ms$,
assuming an initial atom density $n=10^{14}\;\rm atom/cm^{3}$.

Equations (\ref{ALopt}) and (\ref{BLopt}) apply to red as well as blue
detunings so that negative $\Delta-\Delta_e$ gives a
negative $a_{\rm opt}(\Delta,I)$.  Using the parameters in Fig. 2,
it should be possible to tune the
scattering length $a_{\rm scat}$ from $\approx-800\; a_{0}$ to
zero to $\approx+400\; a_{0}$ while the loss rate remains below
$1.7\times10^{-12}\;\rm cm^{3}/s$.

The essential molecular physics that sets the magnitude of $l_{\rm opt}$
depends on the molecular Franck-Condon factor in Eqs.~(\ref{StimGamma})
and (\ref{FCF}).  The transition dipole moment cancels in the definition of
$l_{\rm opt}$ in Eq.~(\ref{Lopt}) due to the relation $\Gamma_{e,\rm nat} \approx
2\Gamma_A$.  However,
the small natural line width of the $^{1}S_{0}$--$^{3}P_{1}$ transition for
alkaline-earth atoms has an important role to play, since much smaller
detuning from atomic resonance can be used than in the case of
alkali-metal atoms.  Both numerical and analytic calculations, similar
to those used in the  Appendix of Ref.~\cite{bohn99}, show that the
magnitude of the Franck-Condon factor (ignoring an oscillating phase factor)
increases as the binding energy of the excited level decreases.
To operate at sufficiently large detuning to satisfy Eqs.~(\ref{critD})
and (\ref{critA}) requires a quite large binding energy for strongly
allowed transitions, i.e, many GHz to more than a THz.  On the other hand,
binding energies in the MHz range can be used in the case of weak
intercombination line transitions.  Consequently we find that optical
lengths can be orders of magnitude larger for weak intercombination
lines than allowed transitions, since the Franck-Condon factors can
be intrinsically much larger once the necessary conditions are satisfied
for changing scattering length without major losses.

In summary, we have shown that molecular energy levels close to the
$^{1}S_{0}+^{3}P_{1}$ dissociation limit of alkaline-earth metal atoms
provide optical Feshbach resonance states that allow for a significant change
of scattering length even for moderate laser intensities and laser
frequencies far detuned from optical resonance, with a
relatively small trap loss rate coefficient.  A small loss rate coefficient
leads to longer observation times. Although we have used
calcium as an example system, we expect our conclusions to remain valid for
other alkaline-earth atoms as well as atoms with similar electronic structure
like ytterbium \cite{takasu03}. Optical Feshbach resonances seem to be a
promising tool that will allow a tunable interaction strength in atomic systems
which do not have magnetic Feshbach resonances.

This work has been partially supported by the U.S. Office of Naval
Research. The research is part of the program of the National
Laboratory FAMO in Toru\'n, Poland.

\end{document}